\documentclass[global,twocolumn]{svjour}
\usepackage{latexsym}
\usepackage{graphics}
\usepackage{amsmath}
\usepackage{amssymb}
\usepackage{mathrsfs}
\usepackage{color}

\newcommand{\JJJ}{\boldsymbol J}
\newcommand{\EEE}{\boldsymbol E}
\newcommand{\DDD}{\boldsymbol D}
\newcommand{\kkk}{\boldsymbol \kappa}
\newcommand{\kfs}{\boldsymbol k}
\newcommand{\nablabf}{\boldsymbol \nabla}
\newcommand{\rrr}{\boldsymbol r}
\newcommand{\aaa}{\boldsymbol \alpha}
\newcommand{\RRR}{\boldsymbol R}
\newcommand{\vvv}{\boldsymbol v}
\newcommand{\highlight}{\color{black}}

\journalname{Microfluidics and Nanofluidics}
\begin{document}
\title{Liquid-infiltrated photonic crystals\thanks{Invited paper for the "Optofluidics" special issue edited by Prof. David Erickson.}}
\subtitle{--- enhanced light-matter interactions for lab-on-a-chip
applications}

\author{Niels Asger Mortensen \and Sanshui Xiao \and Jesper Pedersen}%

\institute{MIC -- Department of Micro and Nanotechnology,
Nano$\bullet$DTU, Technical University of Denmark, {\O}rsteds
Plads, DTU-building 345 east, DK-2800 Kongens Lyngby, Denmark. }
\date{Received: May 11, 2007 / Revised version: June 20, 2007}
\maketitle
\begin{abstract}
Optical techniques are finding widespread use in analytical
chemistry for chemical and bio-chemical analysis. During the past
decade, there has been an increasing emphasis on miniaturization
of chemical analysis systems and naturally this has stimulated a
large effort in integrating microfluidics and optics in
lab-on-a-chip microsystems. This development is partly defining
the emerging field of optofluidics. Scaling analysis and
experiments have demonstrated the advantage of micro-scale devices
over their macroscopic counterparts for a number of chemical
applications. However, from an optical point of view, miniaturized
devices suffer dramatically from the reduced optical path compared
to macroscale experiments, e.g. in a cuvette. Obviously, the
reduced optical path complicates the application of optical
techniques in lab-on-a-chip systems. In this paper we
theoretically discuss how a strongly dispersive photonic crystal
environment may be used to enhance the light-matter interactions,
thus potentially compensating for the reduced optical path in
lab-on-a-chip systems. Combining electromagnetic perturbation
theory with full-wave electromagnetic simulations we address the
prospects for achieving slow-light enhancement of
Beer--Lambert--Bouguer absorption, photonic band-gap based
refractometry, and high-$Q$ cavity sensing.
\end{abstract}
\section{Introduction}\label{sec:1}

Optofluidics is a new branch within photonics which attempts to
unify concepts from optics and
microfluidics~\cite{Psaltis:2006,Monat:2007a}. The purpose of
having these two fields play in concert is at least twofold;

\begin{itemize}
\item[i)] Optofluidics may provide new means for doing optics,
including fluidic tuning of
optics~\cite{Erickson:2006,Diehl:2006,Levy:2006,GersborgHansen:2007}.

\item[ii)] Optofluidics may find
widespread applications in lab-on-a-chip
systems~\cite{Verpoorte:2003,Mogensen:2004,Balslev:2006,Choi:2006}
and analytical chemistry in general, where optical techniques are
being frequently used for chemical and bio-chemical
analysis~\cite{Skoog:1997}.

\end{itemize}

\begin{figure}[b!]
\resizebox{\columnwidth}{!}{
\includegraphics{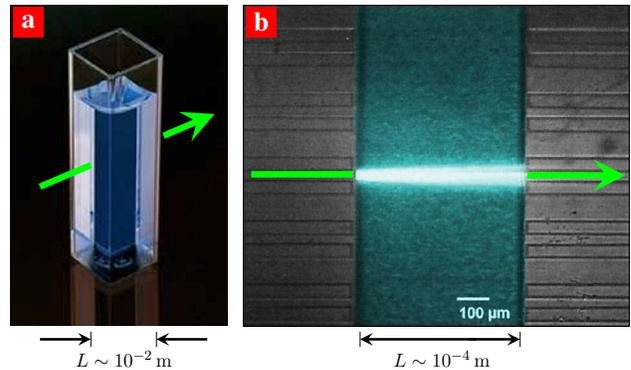}
}
\caption{Panel (a) illustrates a typical macroscopic cuvette while
panel (b) shows a microscope image (top-view) of an equivalent
lab-on-a-chip implementation of a microfluidic channel (vertical
direction) integrated with planar optical waveguides (horizontal
direction). Courtesy of K.~B. Mogensen and J.~P. Kutter (MIC --
Department of Micro and Nanotechnology, Technical University of
Denmark, www.mic.dtu.dk/microTAS).} \label{fig1}
\end{figure}

The increasing awareness from the lab-on-a-chip community of
optofluidics is to a large extend driven by the past decade's
emphasis on miniaturization of chemical analysis
systems~\cite{Janasek:2006} which has naturally stimulated a large
effort in integrating
microfluidics~\cite{Squires:05,Whitesides:2006} and optics in
lab-on-a-chip microsystems~\cite{Verpoorte:2003,Mogensen:2004}.
While a broad variety of phenomena, properties, and applications
benefit from the down-scaling in size~\cite{Janasek:2006}, the
conditions for optical sensing and detection are impeded
tremendously by the reduced optical path length in lab-on-a-chip
systems. Panel (b) in Fig.~\ref{fig1} illustrates a typical
lab-on-a-chip implementation of an absorbance cell. The optical
path length $L$ is often reduced by several orders of magnitude
compared to typical macroscopic counterparts, such as the cuvette
shown in panel (a). A typical size reduction by two orders of
magnitude will penalize the optical sensitivity in an inversely
proportional manner, e.g. in a quantitative concentration
measurement the lowest detectable chemical concentration will be
increased by the same order of magnitude~\cite{Mogensen:2003}.

The above problem may conveniently be explained by the reduced
light-matter interaction time $\tau$ given by the Wigner--Smith
delay time. For a homogeneous system of length $L$ we have
\begin{equation}\label{eq:tau}
\tau\sim L/c,
\end{equation}
with $c$ being the speed of light. For small systems the photons
will thus eventually have too little time for interacting with the
chemical species and only very high concentrations can be
quantified. Many applications involve {\it i)} minute sample
volumes containing {\it ii)} very low concentrations of e.g.
bio-molecules. There is thus obviously a strong call for
miniaturized systems acknowledging the former without jeopardizing
optical approaches quantifying the latter. In this work we explore
the use of photonic crystal concepts for such a purpose.

\begin{figure}[t!]
\resizebox{\columnwidth}{!}{\includegraphics{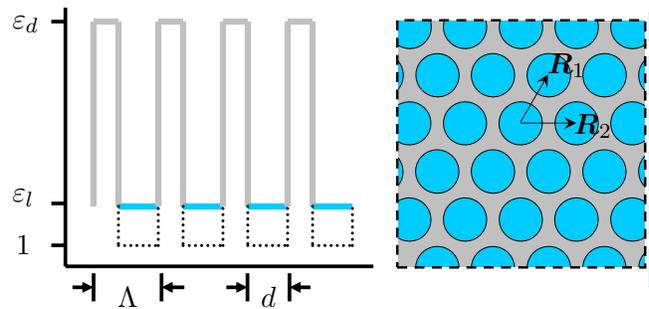}}
\caption{Schematic of the dielectric function variations in a
photonic crystal. High-index material with dielectric function
$\varepsilon_d$ is shown in gray while the liquid with dielectric
function $\varepsilon_l$ is indicated by blue coloring. The dashed
curve indicates the dielectric function variation in the absence
of liquid.} \label{fig2}
\end{figure}

Photonic crystals~\cite{Yablonovitch:1987,John:1987} constitute a
class of artificial electromagnetic structures offering highly
engineered dispersion properties not offered to us by any elements
of Nature herself. Photonic crystals are porous structures, but
with the voids arranged in a highly regular way, see
Fig.~\ref{fig2}). This makes electromagnetic radiation interact
strongly with the matter for wavelengths $\lambda$ comparable to
the periodicity $\Lambda$ of the photonic crystal. Photonic
crystals are strongly dispersive environments supporting a number
of phenomena including photonic band gaps and slow-light
propagation~\cite{Joannopoulos:1995,Sakoda:2005}. As we shall see,
liquid-infiltrated photonic crystals will overall inherit the
unusual dispersion properties from their non-infiltrated
counterparts, thus also changing the way light interacts with
bio-molecules dissolved in the liquid. The effects can be quite
spectacular compared to light-matter interactions in a spatially
homogeneous liquid environment.

Previous experimental work on liquid-infiltrated photonic crystals
may roughly be classified as studies emphasizing either fluidic
tuning of optical properties~\cite{Erickson:2006,Samakkulam:2006}
or work focusing on optical sensing in fluidic
en\-viron\-ments~\cite{Choi:2006,Loncar:2003,Topol'ancik:2003,Chow:2004,Adams:2005a,Chakravarty:2005,Hasek:2006,Skivesen:2007a,Lee:2007}.
On the {\highlight  theoretical} side there has also recently been
a considerable attention to tuning~\cite{Sharkawy:2005} and
sensing
aspects~\cite{Kurt:2005,Kurt:2005a,Prasad:2006,Xiao:2006b,Mortensen:2006d,Mortensen:2007a,Xiao:2007a}
In passing, we note that somewhat parallel to the above
development there has been an effort in fluidic tuning of photonic
crystal fibers~\cite{Mach:2002,Domachuk:2004a,Domachuk:2004b} and
microstructured fibers have recently been integrated in
lab-on-a-chip systems for sensing
applications~\cite{Rindorf:2006}. {\highlight Finally,
infiltrating with liquid crystals~\cite{Yablonovitch:1999} seems
to be an interesting direction for dynamically tunable photonic
crystals~\cite{Busch:1999} and photonic crystal fiber
devices~\cite{Larsen:2003}.}

In this paper we outline our recent theoretical and numerical
work~\cite{Xiao:2006b,Mortensen:2006d,Mortensen:2007a,Xiao:2007a}
on liquid-infiltrated photonic crystals including potential
applications for enhanced light-matter interactions in
lab-on-a-chip systems. The remaining parts of the paper are
organized as follows: in Sec.~\ref{sec:2} we introduce our
theoretical formalism, in Sec.~\ref{sec:3} we address the
dielectric properties of physiological liquids, in
Sec.~\ref{sec:4} we discuss general aspects of modes in
liquid-infiltrated photonic crystals, and in Sec.~\ref{sec:5} we
give examples of potential applications in optical sensing.
Finally, in Sec.~\ref{sec:6} a discussion and conclusions are
given.

\section{Formalism}\label{sec:2}

We consider photonic crystals made from a single high-index
material (with dielectric function $\varepsilon_d$) which can be
of either void-like or pillar-like nature. By infiltrating the
photonic crystal by a liquid (with dielectric function
$\varepsilon_l=n_l^2$) we have a spatially dependent dielectric
function
\begin{equation}\label{eq:epsilon}
\varepsilon(\rrr)=\left\{\begin{matrix}\varepsilon_d &,& \rrr\in
{\mathscr V}_{d},\\\\ \varepsilon_l&,&\rrr \in {\mathscr V}_{l},
\end{matrix}\right.
\end{equation}
where ${\mathscr V}_{d}$ and ${\mathscr V}_{l}$ are the domains of
dielectric material and liquid, respectively. Fig.~\ref{fig2}
illustrates an example of a planar photonic crystal and also
indicates how the dielectric function changes upon liquid
infiltration.

\subsection{The wave equation}
For the dielectric problem constituted by Eq.~(\ref{eq:epsilon}),
the Maxwell equations lead to the following generalized eigenvalue
problem for an electrical field with a harmonic time
dependence~\cite{Joannopoulos:1995,Sakoda:2005}
\begin{equation}\label{eq:E}
\nablabf\times\nablabf\times\big|\EEE_m\big>=\varepsilon\frac{\omega_m^2}{c^2}\big|\EEE_m\big>
\end{equation}
with $\omega_m$ being the angular eigenfrequency of the $m$th
eigenfield $\big|\EEE_m\big>$. We have for simplicity suppressed
the spatial dependence of the dielectric function
$\varepsilon(\rrr)$. Since the eigenvalue problem is of the
generalized form, the eigenfields are orthonormal in the sense
that
\begin{equation}\label{eq:orthonormal}
\big<\EEE_n\big|\varepsilon\big|\EEE_m\big>=\delta_{nm}
\end{equation}
where we have used the Dirac bra-ket notation with inner products
between functions $g$ and $h$ defined according to

\begin{equation}\label{eq:innerproduct}
\big<g\big|h\big> = \int d\rrr\, g^* (\rrr)h(\rrr)
\end{equation}
with the integral extending over all space unless otherwise
indicated.

\subsection{Bloch modes of photonic crystals}
For a photonic crystal the dielectric function $\varepsilon$ of
Eq.~(\ref{eq:epsilon}) is periodic in space, i.e.
$\varepsilon(\rrr+\RRR)=\varepsilon(\rrr)$ with $\RRR$ being a
lattice vector, see Fig.~\ref{fig2}. The solutions to
Eq.~(\ref{eq:E}) are in this case known as Bloch
modes~\cite{Joannopoulos:1995,Sakoda:2005} and are characterized
by both a mode index $m$ and a Bloch wave vector $\kkk$.
Introducing the explicit time and space dependencies we have
\begin{equation}
\big|\EEE_{m,\kkk}(\rrr,t)\big>=
\exp\left[i\left(\kkk\cdot\rrr-\omega_m
t\right)\right]\big|\EEE_{m,\kkk}(\rrr)\big>
\end{equation}
where we have introduced the Bloch function
$\big|\EEE_{m,\kkk}(\rrr)\big>$ which inherits the periodicity of
$\varepsilon(\rrr)$. Typically, one has to rely on numerical
approaches to the calculation of the Bloch modes of
Eq.~(\ref{eq:E}) in a photonic crystal. For this purpose we in
this work employ a plane-wave expansion method with periodic
boundary conditions~\cite{Johnson:2001} yielding the dispersion
relation $\omega_m(\kkk)$. In general, the dispersion in a
photonic crystal differs significantly from the free-space
dispersion relation $\omega(\kfs)=ck$.

\subsection{Perturbation theory}
Many aspects of sensing will be adequately addressed with the aid
of perturbation theory and in the following we briefly outline the
basic equations of first-order electromagnetic perturbation
theory. We assume we make a small frequency-independent
perturbation $\Delta\varepsilon$ to the dielectric function, i.e.
\begin{subequations}
\begin{equation}
\varepsilon\rightarrow \varepsilon + \Delta\varepsilon+\ldots
\end{equation}
The perturbation is small in the sense that
$\Delta\varepsilon\ll\varepsilon$ and we would like to know how
this is going to change the eigenfrequencies and eigenfields.
Expanding the frequency and the field in a similar manner
\begin{equation}
\omega\rightarrow \omega+ \Delta\omega + \ldots
\end{equation}
\begin{equation}
\big|\EEE\big>\rightarrow \big|\EEE\big>+ \big|\Delta
\EEE\big>+\ldots
\end{equation}
\end{subequations}
and substituting into the wave equation, Eq.~(\ref{eq:E}), we may
group terms according to the order of the correction. To lowest
order in the perturbation we arrive at the standard result that
fields remain unchanged while the eigenfrequencies shift according
to
\begin{equation}\label{eq:perturbation}
\Delta\omega_m=-\frac{\omega_m}{2}
\frac{\big<\EEE_m\big|\Delta\varepsilon\big|\EEE_m
\big>}{\big<\EEE_m\big|\varepsilon\big|\EEE_m\big>}.
\end{equation}
For the applications that we have in mind the perturbation will
typically be caused by chemical changes or reactions in the
liquid, i.e. the perturbation will be non-zero in the liquid part
of space only. Thus, in general we will encounter problems with
$\Delta\omega_m\propto f_m$ where
\begin{equation}\label{eq:domega}
f_m\equiv
\frac{\big<\EEE_m\big|\varepsilon\big|\EEE_m\big>_{{\mathscr
V}_l}}{\big<\EEE_m\big|\varepsilon\big|\EEE_m\big>_{{\mathscr
V}_{l+d}}} =\frac{\big<\EEE_m\big|\DDD_m\big>_{{\mathscr
V}_l}}{\big<\EEE_m\big|\DDD_m\big>_{{\mathscr V}_{l+d}}}
\end{equation}
is the filling fraction quantifying the relative optical overlap
with the liquid, i.e. $0\leq f\leq 1$. In the second equality of
Eq.~(\ref{eq:domega}) we have introduced the displacement field
$\big|\DDD\big>=\varepsilon\big|\EEE\big>$. The subscripts on the
inner products emphasize the spatial domain of integration in
Eq.~(\ref{eq:innerproduct}). Note that compared to the definition
normally used by the photonic crystal
community~\cite{Joannopoulos:1995} our definition quantifies the
relative overlap with the liquid-part of space rather than the
dielectric part given by $1-f$ as easily seen from
Eq.~(\ref{eq:orthonormal}).

For a fixed frequency the result of the perturbation may
alternatively be viewed as a shift $\Delta\kkk$ in the Bloch wave
vector $\kkk$. Mathematically, the shift is calculated via the
chain-rule
\begin{equation}\label{eq:deltakappa}
\Delta\kkk_m =
\left(\frac{\partial\omega_m}{\partial\kkk}\right)^{-1}\Delta\omega_m=
\frac{\omega_m}{2}
\frac{\Delta\varepsilon_l}{\varepsilon_l}\frac{f_m}{\vvv_{g,m}}
\end{equation}
where $\vvv_{g,m}=\partial\omega_m/\partial\kkk$ is the group
velocity of the $m$th mode.

In general, for enhanced light-matter interaction phenomena there
is a balance between {\it i)} modifying the dispersion (compared
to propagation through the homogeneous liquid) and {\it ii)} still
maintaining a reasonable optical overlap with the liquid, i.e.
keeping $f$ of the order unity rather than close to zero.

\section{Dielectric properties of physiological liquids}
\label{sec:3}

In this work we emphasize applications involving physiological
liquids as can be found in biological systems. At low frequencies
water is highly polarizable because of the polar nature of the
H$_2$O molecules. In the static limit the relative dielectric
function is around $\varepsilon\sim 80$. However, at optical
frequencies the polar molecules move too slowly in respond to the
temporally rapidly varying field and water becomes transparent to
visible and infrared radiation. The dielectric function is
typically of the order $\varepsilon\sim (1.33)^2\simeq 1.8$.
Compared to highly purified distilled water, physiological liquids
have natural contents of various salts in dissolved form. One
example could be potassium chloride (KCl) which has a high
solubility in water. The potassium (K$^+$) and chlorine (Cl$^-$)
ions are highly mobile in water with a mobility around $\mu_{\rm
ion}\sim 10^{-7}\,{\rm m}^2 ({\rm V\, s})^{-1}$. For ions carrying
a charge $Ze$ the ionic conductivity is given by
\begin{equation}
\sigma = Ze C_{\rm ion} \mu_{\rm ion}
\end{equation}
where $c_{\rm ion}$ is the ionic concentration. Typically, the
ionic concentration is of the order $c_{\rm ion}\sim 1\,{\rm mM}$
yielding conductivities of the order $\sigma_{\rm ion}\sim
10^{-3}\,{\rm S/m}$~\cite{Bruus:2007}. The electrical field will
induce a current density $\JJJ=\sigma\EEE$ in the liquid, but from
an optical point of view the form of Eq.~(\ref{eq:E}) is formally
retained, but now with a complex valued dielectric function
\begin{equation}\label{eq:epsiloncomplex}
\varepsilon_l\rightarrow \varepsilon_l + i\frac{\sigma}{\omega}.
\end{equation}
Consequences of the imaginary part were discussed in
Ref.~\cite{Mortensen:2006d}. We here emphasize the quite similar
form of Eq.~(\ref{eq:epsiloncomplex}) to the Drude model often
employed for the response of metals at optical frequencies. For a
liquid it will be an adequate description of its bulk properties,
thus neglecting any surface chemistry as well as Debye-layer ion
accumulation at the interfaces to other materials such as the
high-index material forming the photonic crystal. Another issue
could be electro-hydrodynamic phenomena where momentum is
transferred from the ionic motion to the fluid (see e.g.
Ref.~\cite{Mortensen:2005} and references therein), but at optical
frequencies such phenomena are strongly suppressed, since the
Debye screening layer forms too slowly in response to the rapidly
varying electrical field. It is common to introduce the Debye
response time
\begin{equation}
\tau_D=\omega_D^{-1}=\varepsilon_l/\sigma
\end{equation}
and for typical electrolytes, $\tau_D$ is of the order a micro
second corresponding to a Debye frequency in the megahertz regime.
For optical frequencies, $\omega\gg \omega_D$, it is thus fully
adequate to treat the Ohmic damping due to the imaginary part in
Eq.~(\ref{eq:epsiloncomplex}) perturbatively. The damping is given
by $\aaa_m=2{\rm Im}\{\kkk_m\}$ which from
Eq.~(\ref{eq:deltakappa}) becomes
\begin{equation}\label{eq:alpha}
\aaa_m^{\rm ion}=f_m\times \frac{\omega_D}{v_{g,m}}.
\end{equation}
The attenuation will thus quite intuitively increase with a
slowing down of the electromagnetic mode near photonic-band edges.
Modes with a large optical overlap with the liquid of course
suffer most from Ohmic dissipation as reflected by the
proportionality of $\aaa$ to $f$. In a spatially homogenous
liquid, Eq.~(\ref{eq:alpha}) with $f=1$ results in a damping of
the order $\omega_D/c$ corresponding to a typical decay length of
the order 1~m. For applications of liquid-infiltrated photonic
crystals in lab-on-a-chip applications it thus seems adequate to
treat typical physiological liquids and the high-index dielectric
material at an equal footing, {\it i.e.} both materials are to a
first approximation assumed lossless.

\begin{figure}[t!]
\resizebox{\columnwidth}{!}{\includegraphics{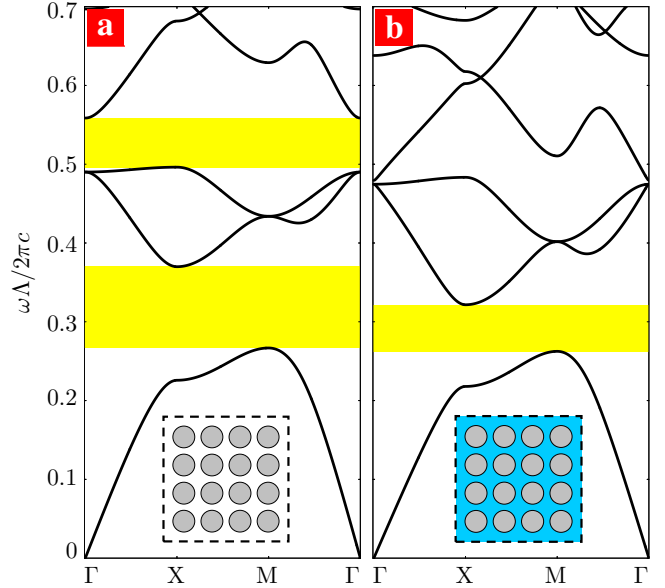}}
\caption{TM modes for a square-lattice of pillars with diameter
$d/\Lambda=0.5$ and dielectric function $\varepsilon_d=10.5$.
Panel (a) shows results for the non-infiltrated photonic crystal
and panel (b) is for the same photonic crystal infiltrated by a
liquid with dielectric function $\varepsilon_l=(1.33)^2$.}
\label{fig3}
\end{figure}

\begin{figure}[b!]
\resizebox{\columnwidth}{!}{\includegraphics{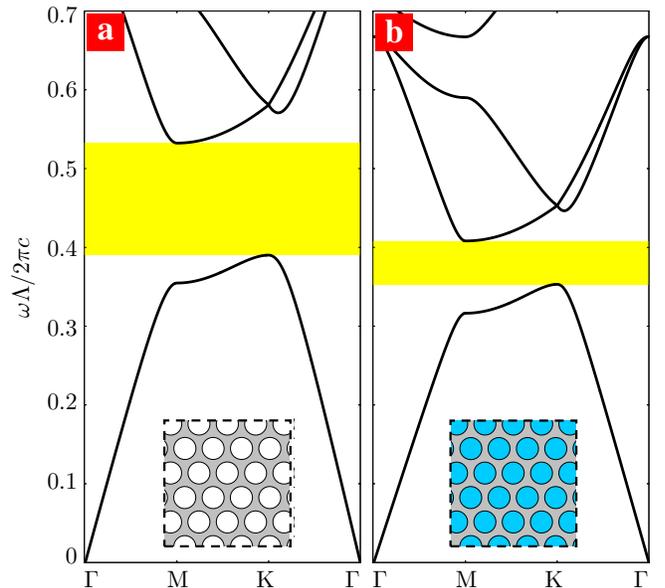}}
\caption{TE modes for a triangular-lattice of voids with diameter
$d/\Lambda=0.96$ in a dielectric material with
$\varepsilon_d=10.5$. Panel (a) shows results for the
non-infiltrated photonic crystal and panel (b) is for the same
photonic crystal infiltration by a liquid with dielectric function
$\varepsilon_l=(1.33)^2$.} \label{fig4}
\end{figure}

\begin{figure}[b!]
\resizebox{\columnwidth}{!}{
\includegraphics{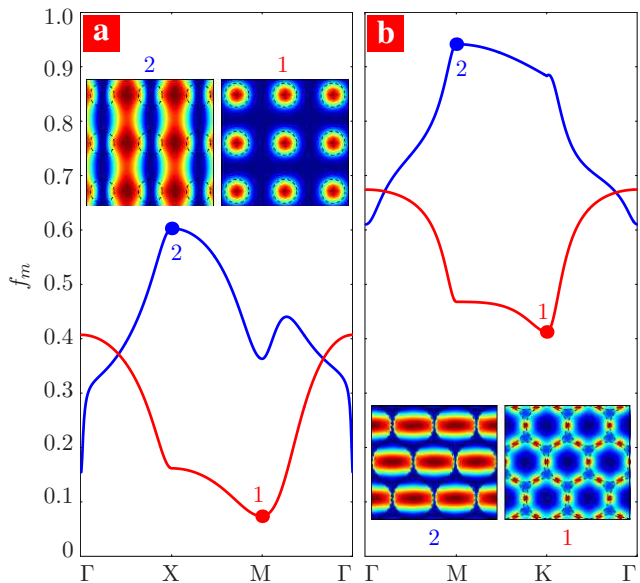}
}
\caption{{\highlight Filling fraction for first dielectric (red
line) and liquid (blue line) bands. Panel (a) shows results
corresponding to the band structure in panel (b) of
Fig.~\ref{fig3} while panel (b) shows results corresponding to
panel (b) of Fig.~\ref{fig4}. The insets show the electrical field
intensity $|\EEE|^2$ at various high-symmetry points for the two
type of bands.}} \label{fig5}
\end{figure}

\section{Classification of photonic bands}
\label{sec:4}

The majority of experimental work on liquid-infiltrated photonic
crystals has so far emphasized void-like structures. The
pillar-like structures may be viewed as the geometrical inverse
with the high and low-index materials being swapped. From
Babinet's principle~\cite{BornWolf} one might imagine the two
types of photonic crystals to have somewhat related properties.
However, the source of formation of photonic band gaps are quite
different in nature; for void-like structures the band-gaps arise
from multiple reflections, i.e. Bragg scattering while band-gaps
relate to Mie resonances for pillar-like
structures~\cite{Sigalas:1996,Lidorikis:2000}. The sensitivity to
infiltration by liquids is also somewhat different in terms of
polarization. The discussion of dielectric versus air bands in
Ref.~\cite{Joannopoulos:1995} is naturally extended to the present
context where we will refer to the bands as dielectric or liquid
bands. Fig.~\ref{fig3} illustrates the typical TM properties of a
square-lattice pillar-like structure. {\highlight The band
structure has been obtained with the aid of a plane-wave
simulation~\cite{Johnson:2001}.} The lowest band is obviously a
dielectric band with $f\sim 0$ as clearly seen by the modest
change upon infiltration. The second band is on the contrary a
liquid band and the higher value of $f$ causes a pronounced
frequency shift of the upper band-gap edge upon infiltration by
the liquid. Due to the reduced index contrast, especially
high-order band-gaps may close upon liquid infiltration as seen by
comparing panels (a) and (b) in Fig.~\ref{fig3}.

Typical TE properties of a triangular-lattice void-like structure
is illustrated in Fig.~\ref{fig4}. {\highlight The band structure
has been obtained with the aid of a plane-wave
simulation~\cite{Johnson:2001}.} Again, the lowest band is
predominantly a dielectric band with $f\sim 0.4$ as clearly seen
by the modest change upon infiltration, though this is not as
pronounced as for the pillar-like structure in Fig.~\ref{fig3}.
Likewise, the second band is a liquid band with a high value of
$f$ causing a pronounced shift of the upper band-gap edge upon
infiltration by the liquid.

Modes with a liquid-band behavior are of course of primary
interest since {\highlight the high electrical field intensity in
the liquid allows} for optofluidic tuning of the dispersion or
likewise highly sensitive optical monitoring of chemical changes
in the fluid. {\highlight Fig.~\ref{fig5} shows results for the
filling fraction $f$ along with field-intensity plots for the
dielectric and liquid bands associated with the band structures in
panels (b) of Figs.~\ref{fig3} and \ref{fig4}. High values of $f$
correlates with a pronounced field localization in the liquid
while low values is associated with a marked field overlap with
the dielectric.}

\begin{figure}[t!]
\resizebox{\columnwidth}{!}{\includegraphics{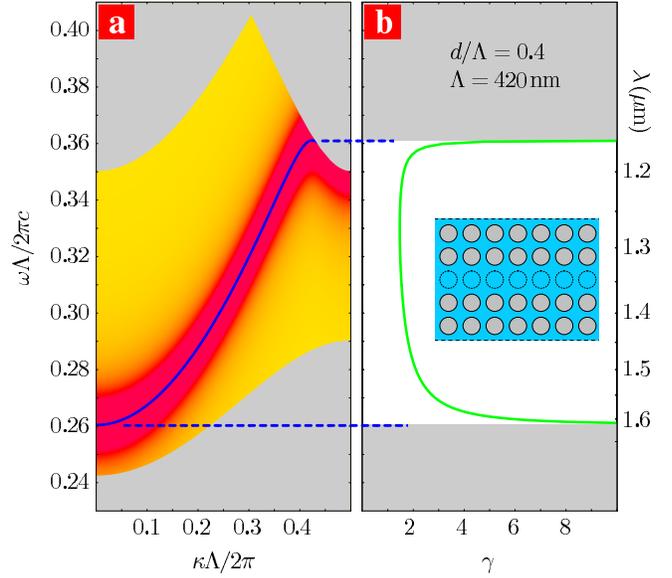}}
\caption{Slow-light enhanced absorbance. Panel (a) shows the band
structure for propagation of TM polarized light along the $\rm
\Gamma$X direction in a line-defect waveguide in a square lattice
of period $\Lambda$ with dielectric rods of diameter
$d=0.4\Lambda$ and $\varepsilon=10.5$ with the structure
infiltrated by a liquid with $\varepsilon=(1.33)^2$. The complete
photonic band gap of the photonic crystal is indicated by yellow
shading while grey shading indicates the finite density-of-states
in the photonic crystal due to the projected bands in the
Brillouin zone. Panel (b) shows the corresponding enhancement
factor $\gamma$ which exceeds unity over the entire bandwidth. The
right $y$-axis shows the results in terms of the free-space
wavelength when results are scaled to a structure with
$\Lambda=420\,{\rm nm}$.} \label{fig6}
\end{figure}

\begin{figure*}[t!]
\begin{center}
\resizebox{0.85\textwidth}{!}{\includegraphics{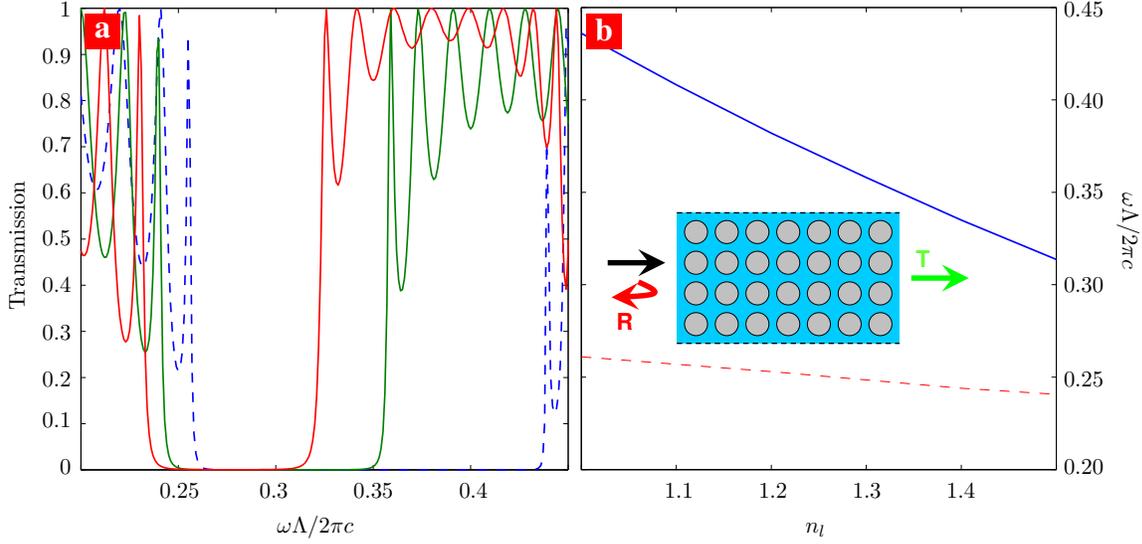}}
\end{center}
\caption{{\highlight Transmission of TM} polarized light incident
on a liquid-infiltrated photonic crystal based on a square-lattice
of pillars with diameter $d/\Lambda=0.4$ and dielectric function
$\varepsilon_d=10.5$. Panel (a) shows the spectral transmission
for the non-infiltrated photonic crystal (dashed line) and when
infiltrated with a liquid with $n_l=1.33$ (solid green line) and
$n_l=1.5$ (solid red line), respectively. Panel (b) shows the
spectral shift of the upper {\highlight (solid line)} and lower
{\highlight (dashed line)} transmission band edges for varying
values of the liquid refractive index $n_l$.} \label{fig7}
\end{figure*}

\section{Applications in sensing and quantitative analysis}
\label{sec:5}

\subsection{Beer--Lambert--Bouguer absorption}

The Beer--Lambert--Bouguer absorbance technique was developed two
centuries ago and has since then become one of the classical
optical workhorses in analytical chemistry~\cite{Skoog:1997}. The
basic principle of an absorbance measurement is illustrated in
Fig.~1. The optical probe, with intensity $I_0$, is incident on a
liquid sample with absorption parameter $\alpha_l$ reflecting a
certain concentration $C$ of some chemical species to be
quantified. Typically the chemicals will be dissolved in a liquid,
but for certain applications gas and even transparent solid phases
are in principle also possible. To lowest order in the
concentration we have
\begin{equation}\label{eq:extinction}
\alpha_l=\sum_i\eta_i C_i,
\end{equation}
with the $C_i$ being the concentration of the absorbing
bio-molecules of the $i$th type with molar extinction coefficient
$\eta_i$.

If we neglect coupling issues, the transmitted intensity $I$ will
then, quite intuitively, be exponentially damped,
\begin{equation}\label{eq:I}
I=I_0\exp(-\gamma \alpha_l L).
\end{equation}
We have, by hand, introduced $\gamma$ as a dimensionless measure
of possible dispersion enhanced light-matter interactions. For a
uniform medium we of course have $\gamma\equiv 1$ and the
expression then often goes under the name of Beer's law. For a
dilute solution $\alpha_l$ correlates linearly with the
concentration of the absorbing chemical species and Beer's law
then provides simple optical means for detecting and quantifying
the concentration of chemical solutions~\cite{Skoog:1997}.
Evidently, the effect relies heavily on having a sufficiently long
optical path length and the longer $L$ is the lower a
concentration can be monitored for a given sensitivity of the
optical equipment measuring $I/I_0$. Quite obviously,
lab-on-a-chip implementations of chemical absorbance cells are
facing major challenges due to the miniaturization as exemplified
by recent experimental work by Mogensen {\it et
al.}~\cite{Mogensen:2003}. Here, we outline our recent suggestion
to use slow-light phenomena to compensate for the reduced optical
path length~\cite{Mortensen:2007a}. In order to rigorously derive
the enhancement factor $\gamma$ for a liquid-infiltrated photonic
crystal we apply perturbation theory, Eq.~(\ref{eq:deltakappa}),
to the problem of a small imaginary part $i\Delta\varepsilon_l$ in
the liquid. Comparing the corresponding damping $\aaa_m=2{\rm
Im}\{\kkk_m\}$ to the damping in a spatially homogenous liquid,
obtained from Eq.~(\ref{eq:deltakappa}) with $f=1$ and
$v_g=c/n_l$, we arrive at an enhancement factor $\gamma_m \equiv
\alpha_m/\alpha_l$ given by
\begin{equation}
\label{eq:gamma}
\gamma_m =f_m\times\frac{c/n_l}{v_{g,m}}.
\end{equation}
The interpretation of this result is straightforward; slow-light
propagation, $v_g\ll c$, will significantly enhance the
absorbance. The presence of the filling factor $f$ is also quite
obvious since only the relative fraction of the intensity residing
in the liquid will contribute to the absorption. Finally, we note
that for a spatially homogeneous liquid we get $\gamma=1$ as
required by the definition in Eq.~(\ref{eq:I}). In general, the
enhancement factor has a lower bound, $\gamma\geq 0$, while there
is in principal no upper bound since the inverse group velocity
diverges for Bloch wave vectors near the Brillouin zone (such as
the $\rm X$ or $\rm M$ points) or near the center of the zone (the
$\rm \Gamma$ point), see Fig.~\ref{fig3}. In Fig.~\ref{fig6} we
show results for a line-defect waveguide in a square lattice of
period $\Lambda$ with dielectric rods of diameter $d=0.4\Lambda$
and $\varepsilon=10.5$. As expected, the pillar-like structure
supports a mode with a very high optical overlap with the liquid,
$f\sim 1$, combined with a reduced group velocity over the entire
band width of the waveguide transmission window, see panel (a).
Furthermore, at the $\rm \Gamma$ point the inverse group-velocity
diverges {\highlight which is also the case near the upper
waveguide transmission edge. In fact, the enhancement factor is of
the order or larger than 2 over the entire bandwidth. Enhancement
by an order of magnitude is easily achieved near the transmission
band edges.}

\begin{figure*}[t!]
\begin{center}
\resizebox{0.85\textwidth}{!}{\includegraphics{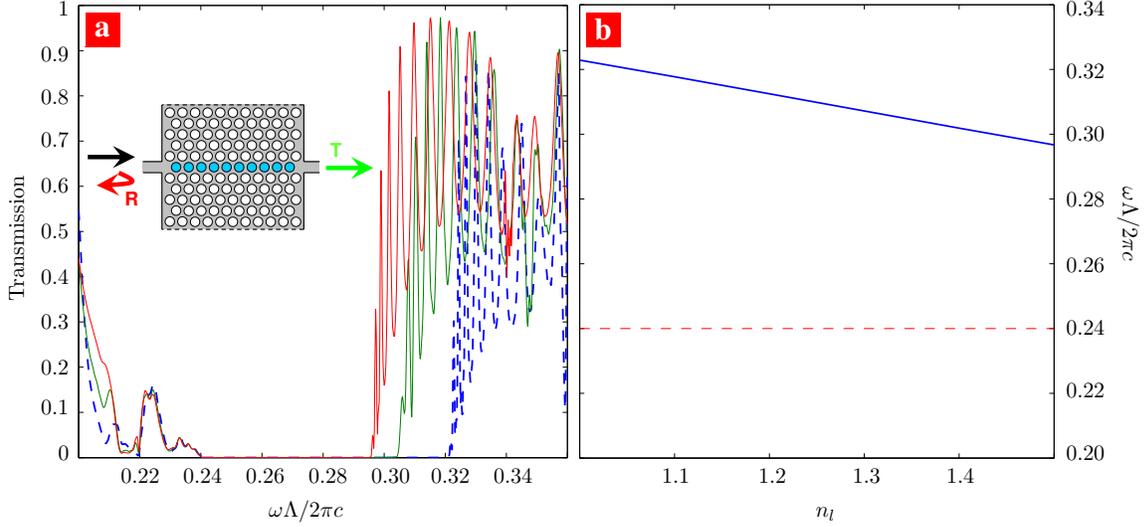}}
\end{center}
\caption{{\highlight Transmission of TE polarized} light incident
on a liquid-infiltrated photonic crystals wave-guide structure
based on a triangular-lattice of voids with diameter
$d/\Lambda=0.6$ and dielectric function $\varepsilon_d=10.5$.
Panel (a) shows the spectral transmission for the non-infiltrated
photonic crystal (dashed line) and when infiltrated with a liquid
with $n_l=1.33$ (solid green line) and $n_l=1.5$ (solid red line),
respectively. Panel (b) shows the spectral shift of the upper
{\highlight (solid line)} and lower {\highlight (dashed line)}
transmission band edges for varying values of the liquid
refractive index $n_l$.} \label{fig8}
\end{figure*}

\subsection{Refractometry}
Refractometry in its basic form is another widely used method
which most often is a qualitative measurement indicating chemical
changes contrary to e.g. absorbance measurements which are of
quantitative nature. Chemical reactions and changes most often
cause a modification $\Delta n_l$ of the refractive index, thus
altering the optical response of e.g. a reflection or transmission
measurement. Chemical changes may thus be monitored by watching
the spectral shifts of e.g. transmission
resonances~\cite{Xiao:2006b,Xiao:2007a}. Obviously, the larger a
spectral shift the more sensitive is the method. As for the
previous problem we will apply perturbation theory to study the
spectral sensitivity due to a small perturbation
$\Delta\varepsilon_l \simeq 2n_l\Delta n_l$. {\highlight From
Eq.~(\ref{eq:perturbation}) we get}
\begin{equation}
\frac{\Delta\omega_m}{\omega_m}=- f_m\times \frac{\Delta
\varepsilon_l}{2\varepsilon_l}
\end{equation}
or introducing the free-space wavelength $\lambda=2\pi c/\omega$
we may express the sensitivity in the form of a wavelength shift
$\Delta\lambda$ due to a shift in liquid refractive index $\Delta
n_l$,
\begin{equation}\label{eq:wavelengthshift}
\frac{\Delta\lambda_m}{\Delta n_l}=f_m\times
\frac{\lambda_m}{n_l}.
\end{equation}
The interpretation is again quite intuitive. In a homogeneous
liquid we have a dispersion $1/\lambda\propto\omega(\kkk) =
(c/n_l)k$ so that $\Delta\lambda/\Delta n_l= \lambda/n_l$ which is
obviously the ultimate sensitivity one can have. However,
transmission {\highlight through homogenous media} is not
associated with any pronounced resonances or cut-offs.
Transmission experiments {\highlight reflect the photonic}
density-of-states and strong resonances may be achieved by
infiltrating a photonic crystal with the
liquid~\cite{Mortensen:2006d}. In that case the sensitivity is
naturally corrected by the filling factor $f$ as shown by the
rigorous perturbative analysis in Eq.~(\ref{eq:wavelengthshift}).
From Eq.~(\ref{eq:wavelengthshift}) the working wavelength is of
course important to the sensitivity. However, this is a somewhat
trivial effect for refractometry in general, while the dependence
on filling fraction in Eq.~(\ref{eq:wavelengthshift}) is really
the limiting factor. The performance of different structures,
ranging from evanescent field sensing devices to
liquid-infiltrated photonic crystals, may thus be compared by
comparing the corresponding filling fractions. Such relative
comparisons has previously been suggested in the
literature~\cite{Kurt:2005}, but Eq.~(\ref{eq:wavelengthshift})
does in fact allow for an absolute definition of the sensitivity.

Evanescent field-sensing devices are characterized by an almost
vanishing $f$ which will typically only be of the order a few
percent. Depending on the design, void-like photonic crystals
offer {\highlight moderate-to-high filling fractions depending on
the void-to-solid volume ratio, and similarly for pillar-like
photonic crystals.}

Figure~\ref{fig7} shows a proposal of refractometry with a
liquid-infiltrated photonic crystal. The simple structure consists
of a square-lattice of pillars with diameter $d/\Lambda=0.4$ and
dielectric function $\varepsilon_d=10.5$. Transmission spectra for
TM polarized light have been obtained by means of a
two-dimensional finite-difference time-domain (FDTD) method, see
Ref.~\cite{Xiao:2006b} and references therein. Panel (a) shows the
spectral transmission for the non-infiltrated photonic crystal
(dashed line) and when infiltrated with a liquid with $n_l=1.33$
(solid green line) and $n_l=1.5$ (solid red line), respectively.
From the band structure in Fig.~\ref{fig3} one would expect a
suppression of the transmission for frequencies inside the
photonic band gap which correlates nicely with the observed
transmission. Furthermore, when infiltrating the photonic crystal
with liquid, the spectral position of the upper transmission band
edge is expected to be more sensitive to the perturbation than the
lower transmission band edge. Panel (b) shows the spectral shift
of the upper and lower transmission band edges for varying values
of the liquid refractive index $n_l$. The upper curve does indeed
have a larger slope than the lower curve which according to
Eq.~(\ref{eq:wavelengthshift}) supports the picture of
transmission through a liquid band and dielectric band,
respectively. The transmission results agrees quantitatively with
the band-structure theory as studied in detail in
Ref.~\cite{Xiao:2006b} and from Eq.~(\ref{eq:wavelengthshift}) we
estimate that $f\sim 80\%$ for the transmission through the liquid
band.

In Fig.~\ref{fig8} we show another implementation of refractometry
employing a liquid-infil\-tra\-ted photonic crystals wave-guide
structure based on a trian\-gular-lattice of voids with diameter
$d/\Lambda=0.6$ and dielectric function $\varepsilon_d=10.5$.
Selective filling of the defect wave-guide structure is of course
crucial, but nanofluidic tuning has recently been demonstrated
experimentally in a similar structure~\cite{Erickson:2006}. Panel
(a) shows the spectral transmission of TE polarized light for the
non-infil\-tra\-ted photonic crystal (dashed line) and when
infiltrated with a liquid with $n_l=1.33$ (solid green line) and
$n_l=1.5$ (solid red line), respectively. Panel (b) shows the
spectral shift of the upper and lower transmission band edges for
varying values of the liquid refractive index $n_l$. Again, the
upper curve corresponds to transmission through a band of
predominantly liquid-band nature with $f\sim 40\%$ estimated from
Eq.~(\ref{eq:wavelengthshift}).

{\highlight Above we have emphasized transmission properties, but
spectral shifts could alternatively be monitored in a reflection
configuration.}

\subsection{Cavity based sensing}

Obviously, cavity modes may be used for refractometry where
Eq.~(\ref{eq:wavelengthshift}) also applies. Defect structures in
photonic crystals are quite interesting in this context and in
recent experiments Chow {\it et al.} have pursued this
direction~\cite{Chow:2004}. For refractometric applications the
quality factor $Q$ of a resonance should be sufficiently high to
easily monitor and resolve even small spectral shifts of the
resonance. Here, we would like to emphasize another potential
usage of high-$Q$ resonators somewhat resembling the
Beer--Lambert--Bouguer absorption technique. For the slow-light
enhancement we note that the divergence of the inverse group
velocity near the edge of the Brillouin zone is basically a
standing wave phenomenon. Thus, cavity modes may in a similar way
be used to enhance the light-matter interaction time $\tau$, but
contrary to the case of Eq.~(\ref{eq:tau}) the Wigner--Smith delay
time now relates to the cavity quality factor $Q$. In the
following we consider an isolated Lorentzian resonance,
\begin{equation}\label{eq:rho}
\rho_m(\omega)=-\frac{1}{\pi}{\rm Im}\left\{
\frac{1}{\omega-\omega_m + i\Gamma_m/2}\right\},
\end{equation}
with the line width $\Gamma_m$ corresponding to an intrinsic
quality factor $Q_{\rm in} = \omega_m/\Gamma_m$. We imagine that
the intrinsic quality factor includes everything but the
broadening due to absorption caused by the presence of
bio-molecules in the liquid. The total quality factor $Q$ may then
be written as
\begin{equation}\label{eq:Q}
Q=\left(Q_{\rm in}^{-1}+Q_{\rm abs}^{-1}\right)^{-1}\simeq
\left\{\begin{matrix}Q_{\rm abs}&,& Q_{\rm abs}\ll Q_{\rm in},\\
\\Q_{\rm in}&,& \alpha_l\rightarrow 0,
\end{matrix}\right.
\end{equation}
where $Q_{\rm abs}$ is the contribution from absorption. High-$Q$
cavities would be of prime interest to sensing since the line
width would be dominated by the presence of absorbing
bio-molecules rather than the intrinsic line width associated with
the localization of light. Obviously, the higher a $Q_{\rm in}$
the lower a concentration of molecules can be detected. In order
to quantify this further we consider the limit of weak absorption
in the liquid and apply perturbation theory. Calculating the
imaginary shift of $\omega_m$ due to
$\Delta\varepsilon=i\Delta\varepsilon_l$ {\highlight from
Eq.~(\ref{eq:perturbation})} and substituting into
Eq.~(\ref{eq:rho}) we identify an additional broadening
corresponding to
\begin{equation}\label{eq:Qabs}
Q_{\rm abs}= f_m^{-1}\times \frac{2\pi n_l}{\alpha_l \lambda_m}
\end{equation}
where we have introduced the absorption parameter $\alpha_l$ of
the liquid. The form of Eq.~(\ref{eq:Qabs}) resembles the form
derived for e.g. homogeneous micro-sphere
resonators~\cite{Gorodetsky:1996} except for the presence of the
filling factor $f$. Obviously, the lower an overlap with the
absorbing liquid the smaller an additional broadening of the
intrinsic resonance line width.

If we consider just one type of molecules, see
Eq.~(\ref{eq:extinction}), the criterion in Eq.~(\ref{eq:Q}) means
that the lowest detectable concentration is bound by
\begin{equation}
C \gtrsim f_m^{-1} \times \frac{1}{\eta\lambda_m Q_{\rm in}}
\end{equation}
with $\eta$ being the molar extinction coefficient. {\highlight
As an example, for a typical value $\eta\sim 1\,{\rm M}^{-1}\,{\rm
\mu m}^{-1}$ a detection in the micro-molar regime would require
$f\times Q_{\rm in}\gtrsim 10^5$.}

Cavity designs optimizing the product $f\times Q_{\rm in}$ are
obviously needed, but it is also well-known that for slab-like
photonic crystals the typical high-$Q$ cavities will have a quite
modest $f$ to prevent degradation of $Q$ by out-of-plane
leakage~\cite{Yoshie:2001,Akahane:2003} though some designs may
get around this~\cite{Asano:2006,Tomljenovic-Hanic:2006}. Indeed,
the refractometric measurements in Ref.~\cite{Chow:2004} suggest
that $f\sim 20\%$ for their cavity mode with $Q\sim 400$. Cavities
in three-dimensional photonic crystals with photonic-band gap
confinement in all three directions could be an interesting
scenario. As an alternative, recent experiments with liquid
micro-droplets~\cite{Azzouz:2006,HosseinZadeh:2006} facilitate
whispering-gallery modes with $f \simeq 1$ and potentially the
intrinsic $Q$-factor can be made very high due to the absence of
surface roughness at the almost perfect liquid-air interface which
is smooth on the molecular length scale.
Ref.~\cite{HosseinZadeh:2006} reported $Q\sim 10^5$ in a
fiber-coupled experiment. Experimentally one might use the above
scheme to quantify concentrations e.g. by time resolved ring-down
spectroscopy or by other means such as transmission experiments
typically employed for characterization of high-$Q$
resonators~\cite{Vahala:2003}.

\section{Conclusion}
\label{sec:6}

With the increasing emphasis on miniaturization of chemical
analysis systems there is presently a strong effort devoted to
integrating microfluidics and optics in lab-on-a-chip
microsystems. We have outlined possible applications of
optofluidics in this context and used simple theoretical
approaches to illustrate how light-matter interactions may be
enhanced in liquid-infiltrated photonic crystals supporting highly
dispersive modes and pronounced resonances.

Our formalism combines electromagnetic perturbation theory with
full-wave electromagnetic simulations. The latter is used in
calculating the dispersion properties of ideal (lossless)
liquid-infiltrated photonic crystals while the former is used
systematically to predict the spectral effects and changes caused
by chemical perturbations of the infiltrated liquid.

In this paper we implicitly focused on near-infrared wavelengths
with values for the high-index material resembling silicon.
However, we emphasize that the general concepts apply equally well
to other wavelength regimes such as the visible regime as in
Ref.~\cite{Choi:2006} or even the terahertz regime emphasized in
Refs.~\cite{Hasek:2006,Kurt:2005,Kurt:2005a}. The phenomena that
we have discussed do to some extend rely on a sufficient index
contrast between the liquid and the high-index materials. While
polymers are certainly attractive materials for future low-cost
disposable lab-on-a-chip devices they might not provide a
sufficient index contrast for all of the applications that we have
discussed and the choice of wavelength and materials should of
course reflect the particular applications in mind.

With this theoretical paper we have sought to emphasize the
potential benefits from using liquid-infiltrated photonic crystals
for sensing purposes in lab-on-a-chip systems such as label-free
sensing of bio-molecules. While photonic crystal concepts are
generally considered a pro\-mising route to highly sensitive
devices, our theory clearly demonstrates an {\highlight inevitable
challenge}; the photonic crystal design needs to balance between
having a strong optical proximity from the photonic crystal
structure and at the same time supporting a decent optical overlap
with the liquid. {\highlight On the other hand, the huge design
space offered by photonic crystals seems to allow us to explore
and cope with this challenge.} In conclusion we believe that
photonic crystals may be used to facilitate enhanced light-matter
interactions which may be traded for yet smaller miniaturized
systems or for increased sensitivity of existing devices.

\begin{acknowledgement}
We thank A. Kristensen, M. Gersborg-Hansen, J.~P. Kutter, K.~B.
Mogensen, H. Bruus, and J. L{\ae}gsgaard for discussions. This
work is financially supported by the \emph{Danish Council for
Strategic Research} through the \emph{Strategic Program for Young
Researchers} (grant no: 2117-05-0037).
\end{acknowledgement}

%\bibliographystyle{springer}
%\bibliography{Q:/papers/BibTeX/OFTS}

\end{document}